\documentstyle[aps,multicol,epsf]{revtex}
\newcommand{\bea}{\begin{eqnarray}}
\newcommand{\eea}{\end{eqnarray}}

\begin{document}
\draft
\title{Dislocation Free Island Formation in Heteroepitaxial Growth: 
An Equilibrium Study}

\author{Istv\'an Daruka and Albert-L\'aszl\'o Barab\'asi}

\address{ Department of Physics, University of Notre Dame,
Notre Dame, IN 46556}

\date{\today}

\maketitle

\begin{abstract}

We investigate the equilibrium properties of strained heteroepitaxial
systems, incorporating the formation and the growth of a wetting film,
dislocation free island formation, and ripening. The derived phase
diagram provides a detailed characterization of the possible growth
modes in terms of the island density, equilibrium island size, and
wetting layer thickness.  Comparing our predictions with experimental
results we discuss the growth conditions that can lead to stable
islands as well as ripening.
\end{abstract}


\begin{multicols}{2}
\narrowtext

Heteroepitaxial growth of highly strained structures has gained
interest lately as it offers the possibility to fabricate nanoscale
islands with very narrow size distribution \cite{mrs}.  Typically, $H$
monolayers of atoms are deposited on a substrate, where the substrate
and the deposited film have different equilibrium lattice constants.
For small coverage the experiments document the pseudomorphical formation of a
wetting film, but after the film reaches a certain critical thickness,
$H_c$, dislocation free islands form on the substrate.  Thanks to
their small and uniform size, these islands, coined self-assembling
quantum dots (SAQD), are candidates for three dimensional electron
confinement \cite{mrs}.

The controlled production of SAQDs for both optical and electronic
applications requires a good description of the basic mechanisms
determining the size and the distribution of the islands. However,
such an understanding is hampered by the coexistence of equilibrium
and nonequilibrium effects: while the
experimentally well documented existence of a flux independent
critical wetting film thickness \cite{gera}, $H_c$, is well
described by equilibrium theories of heteroepitaxial growth
\cite{ter1}, the observed flux and temperature dependence of
the island sizes \cite{mrs} provide direct evidence of nonequilibrium
effects  contributing to the island formation process \cite{bara1}. A detailed
theory of SAQD formation, incorporating both nonequilibrium and
equilibrium effects is beyond reach at this point. However, since for
reversible systems nonequilibrium effects represent the path
of the system towards an equilibrium state, an adequate theory of
SAQD formation should  first provide a detailed
description of the equilibrium states supported by the dislocation
free strained system.
An important step in this direction was taken by Shchukin {\it et al.}
\cite{schu}, who found that depending on the material constants and
the misfit one can obtain either stable islands, or ripening  takes
place in the system.  However, by neglecting the existence of the
wetting layer, their study could not predict the actual growth mode,
 nor could provide the island density,
island size, and the wetting layer thickness as a function of the
deposited material,  quantities that can be measured experimentally
with great accuracy \cite{mrs}.

In this paper we investigate the equilibrium properties of strained
heteroepitaxial systems, incorporating the growth of the wetting film,
dislocation free island formation, and ripening. Our results can be
summarized in a phase diagram, that not only predicts the main growth
modes, but also provides a detailed characterization of the possible
phases in terms of the island density, equilibrium island size and
wetting layer thickness.  We find that the stability of the islands
depends very sensitively on the coverage, i.e. the misfit strain and
the coverage have to exceed a critical value for stable islands to
exist, and that for {\it any} misfit there is a second critical coverage
beyond which ripening occurs.  

{\it Model and free energy---} We consider that $H$ monolayers of atom
A with lattice constant $d_A$ are deposited on top of the substrate B
with lattice constant $d_B$, and are allowed to equilibrate.  Due to
the lattice mismatch, $\epsilon =(d_A-d_B)/d_B$, in equilibrium one
expects that a certain fraction of the atoms A forms a wetting film of
$n_1$ monolayers and the rest of the material ($H-n_1$ monolayers)
is  distributed in 3D islands. We consider that the 3D islands
have a pyramidal shape with a fixed aspect ratio, corresponding to a
single minima in the Wulff's plot. Neglecting
evaporation, the deposited material represents a conserved system in
equilibrium with a thermal reservoir, thus the relevant
thermodynamical potential density is the free-energy per atom, $f = u
- Ts$, where $u$ is the internal energy density, $T$ is the
temperature and $s$ is the entropy density of the system. However one
can show that the entropic contribution to $f$ is negligible, thus $f
\approx u$, where \bea u(H, n_1, n_2, \epsilon ) =
E_{ml}(n_1)+n_2E_{isl} +(H-n_1 -n_2)E_{rip}. \label{inte} \eea The
first term provides the contributions of the $n_1$ strained
overlayers, being an integral over the binding and the elastic energy
densities.  The energy density of a uniformly strained layer is given
by $G=C\epsilon^2-\Phi_{AA}$, where $-\Phi_{AA}$ is the energy of an
AA bond and $C$ is a material constant, being a function of the
Young's modulus and the Poisson's ratio \cite{landau}. At the wetting
layer--substrate interface atoms have AB bonds with the substrate with
a binding energy $-\Phi_{AB}$, such that $\Delta
=\Phi_{AA}-\Phi_{AB}<0$ (wetting condition).  However, due to the
short range intermolecular interactions the binding energies of A
atoms close to (but not at) the substrate is  also  modified
\cite{ter1}: as we move away from the substrate, the binding energy
density increases from $-\Phi_{AB}$ (in the first monolayer) to its
asymptotic value $-\Phi_{AA}$.  These intermolecular forces are
responsible for the critical layer thickness larger than one monolayer
in heteroepitaxy \cite{ter1}.  To include this effect we calculate the
total energy stored in the wetting layer as \bea
E_{ml}(n_1)=\int_0^{n_1}dn\bigl\{ G +\Delta\bigl[(\Theta (1-n)+ \Theta
(n-1)e^{-(n-1)/a}\bigr]\bigr\}, \label{emol} \eea where $\Theta (x)= 0
$ if $x<0$ and $\Theta (x)= 1 $ if $x>0$.  The $a=0$ limit corresponds
to the absence of the short range forces.  While (\ref{emol}) provides
a reasonable fit to the result of Ref. \cite{ter1}, the particular
form of (\ref{emol}) does not modify the qualitative behavior of the
free energy provided that the binding energy is strictly monotonous
and bounded as a function of $n$.

The second term in eq. (\ref{inte}) describes the free energy per atom
of the pyramidal islands and the island-island interaction \cite{schu}
\bea E_{isl}=gC{\epsilon}^2-\Phi_{AA} +E_0\Bigl( -{2\over x^2} \ln
e^{1/2}x +{\alpha \over x} +{{\beta (n_2)} \over x^{3/2}} \Bigr),
\label{eisl} \eea where $x=L/L_0$ is the reduced island size, $L_0$
being a material dependent characteristic length \cite{schu}.
Departures from planar geometries can lead to the relaxation of the
strain energy. Thus the strain energy density of the islands (first
term in eq. (\ref{eisl})) is lower than that of the compressed wetting
layer, this reduction being expressed by the form factor $g$ ($0<g<1$)
\cite{schu}.  The second term stands for the binding energy.  The
elastic energy of an edge of length $L$ is proportional to $-L\ln L$
\cite{mar1}, thus the energy density is $\sim -\ln L/L^2$, accounting
for the first of the three terms in the parenthesis.  The interaction
of the homoepitaxial and heteroepitaxial stress fields leads to a
cross term $-\epsilon /L \sim -\epsilon /x$. Furthermore, the facet
energy is proportional to the area of the facet, $L^2$, giving the
energy density as $\sim 1/L \sim 1/x$.  The cross term and the facet
energies are combined in the second term in the parenthesis of
eq. (3), $\alpha /x$, with $\alpha = p(\gamma - \epsilon)$, where $p$
and $\gamma$ are material constants describing the coupling between
the homoepitaxial and heteroepitaxial stress fields (also function of
the island geometry) and the extra surface energy introduced by the
islands, respectively.  Finally, since the stress fields of the
individual islands overlap, there is island-island interaction,
described by the last term in the parenthesis of eq. (\ref{eisl}),
where $\beta (n)=b{\epsilon}^2{n}^{3/2}$ \cite{schu}. This can be
expressed in terms of the average island spacing
$d=1/\sqrt{\rho_{isl}}$ and the reduced island size $x$, giving the
interaction term as $\sim (x/d)^3$, corresponding to the dipole-dipole
interaction between the islands \cite{srolovitz}.  The energy terms
appearing in eq. (3) are scaled by the characteristic energy $E_0$ set
by the edge energy of an island of size $L_0$.  We also scale $C,
\Phi_{AA}$, and $\Phi_{AB}$ by $E_0$, thus the results are independent
of the numerical value of $E_0$.

The total elastic energy density of the ripened islands can be obtained  
from (\ref{eisl}) by taking the limit $x\rightarrow \infty$,
providing
$E_{rip}=gC{\epsilon}^2-\Phi_{AA}$,
which is multiplied 
by the total number of atoms stored in the ripened islands, $(H-n_1-n_2)$.

{\it Phase diagram---} Eqs. (\ref{inte})--(\ref{eisl}) define the free
energy of the wetting film and 3D pyramidal islands, whose minima
determines the equilibrium properties of the system.  Consequently, we
have to minimize $f$ in respect to $n_1$,  $n_2$,
and  $x$. The growth modes (phases) provided
by the minimization process, as a function of the two most relevant
experimental parameters, the amount of the deposited material $H$ and
the misfit $\epsilon$, are summarized in the phase
diagram shown in Fig. 1.  In the following we discuss the
properties of the phases predicted by our analysis.

{\it FM Phase}:
If $H<H_{c_1}(\epsilon)$, the deposited material 
 contributes to the pseudomorphic growth of the wetting film and islands are
absent, reminiscent of the Frank van der Merve (FM) growth mode.
The free energy  has its minima at $n_2 = 0$ and 
$n_1 =H$, indicating that 
the thickness of the wetting layer coincides with the
nominal thickness of the
deposited material, $H$ (see Fig. 2a).
The growth of the wetting layer continues
until 
$H$ reaches a critical value, $H_{c_1}(\epsilon )$, 
the 
phase boundary between the FM and the $R_1$ or SK$_1$ phase.

{\it $R_1$ Phase}: Above $H_{c_1}(\epsilon )$, for $0<\epsilon<\epsilon_1$,
the free energy develops a new
minima at $n_2=0$ and $0<n_1<H$. 
Consequently, after the formation of a wetting layer of $n_1=H_{c_1}(\epsilon )$
monolayers, the excess material ($H-n_1$) contributes
to the formation of ripened islands. The free energy
decreases monotonically for large $x$,
thus there is a tendency to accumulate all available material ($H-H_c(\epsilon )$)
in as large islands
as possible.  
These ripened islands, being infinitely large, have zero
density.

{\it SK$_1$ Phase}:
Above $H_{c_1}(\epsilon )$, for $\epsilon_1<\epsilon<\epsilon_2$,
the free energy develops a new
minima at nonzero $n_1$ and $n_2$, such that $n_1+n_2=H$, i.e. 
the deposited material ($H$) is distributed between
$n_1$ layers, forming the wetting film, and finite islands, whose
total mass is $n_2$ (see Fig. 2a),
similar to the Stranski-Krastanow (SK) growth mode. 
At $H_{c_1}(\epsilon )$ 
the equilibrium island size jumps from zero (in the FM phase)
to some finite $x_0(H, \epsilon )$ value. 
Naturally, within the SK$_1$ phase,  $n_1$, $n_2$, $x_0$, and the island density $\rho$
are continuous functions of $H$ and $\epsilon$.
As Fig. 2a indicates, with increasing $H$,   
$\rho$ increases from zero at $H_c$ to a finite value. 
Due to the island--island interaction, the wetting layer continues to grow
sub-linearly. 

{\it R$_2$ Phase}:
In this phase, the free energy surface has a minima at 
$0<n_1<H$, $0<n_2<H$, such that $H-n_1-n_2>0$, indicating that the 
deposited material is distributed between
a wetting film, finite islands, and ripened islands ($H-n_1-n_2$).
The finite islands formed 
in the SK$_1$ phase is  preserved (see Fig. 2a and b), being stable in
respect to ripening. Thus
finite and ripened islands coexist in the R$_2$ phase. 

{\it VW Phase}: For large misfit ($\epsilon>\epsilon_2$) 
and for small coverages ($H<H_{c_4}(\epsilon )$), 
the free energy has its minima at $n_2 = H$ and
$n_1 =0$, indicating that all the deposited material
is accumulated in finite islands. 
Due to the large misfit, in this phase the wetting
film is absent and the islands form
directly on the substrate, similar to the Volmer-Weber (VW) growth mode.

{\it SK$_2$ phase}: For $\epsilon_2<\epsilon<\epsilon_3$,
increasing $H$, at $H_{c_4}(\epsilon )$, we reach the
SK$_{2}$ phase.
The behavior of the system is
different from the SK$_1$ growth mode: at the $H_{c_4}$ boundary we have islands
formed in the VW mode. As Fig. 2b indicates, in 
the SK$_2$ phase the island density and the island
size remain unchanged, and a wetting film starts forming. This
process continues
until a full monolayer is completed, at which point  we enter the
SK$_1$ phase. In contrast with the SK$_1$ phase, 
in the SK$_2$ phase   
the formation of further 
islands is
suppressed until the one monolayer thick wetting layer is completed.

{\it R$_{3}$ Phase}:
In this phase, present for $\epsilon>\epsilon_3$ and for
$H>H_{c_4}$, the free energy has its 
minima at $n_1=0$ and $0<n_2<H$, indicating the formation of
ripened islands. The formation of
stable islands 
is suppressed, and all the material deposited after $H_{c_3}$ contributes
only to the ripened
islands, coexisting with the stable islands formed in the VW growth mode.
However, in contrast to R$_2$, in R${_3}$ a wetting film is absent.

{\it Comparison with experiments---}
A quantitative comparison of the phase diagram 
with the experiments requires the knowledge of the material constants 
in (\ref{inte})--(\ref{eisl}), which determine
the precise value of $\epsilon_1$, $\epsilon_2$,  $\epsilon_3$,  and 
the location of the lines in the phase diagram.
However, 
the {\it topology} of the phase diagram is {\it material independent}, as
long as the SK phase is supported by the system. This
robustness of the phase diagram implies that in equilibrium these are the {\it only
phases}  supported by  the free energy (\ref{inte})--(\ref{eisl}).

During ripening, when the island size reaches a certain
critical size, dislocation formation relaxes  the strain
energy of the islands, allowing the fast growth of dislocated islands.
Furthermore, many  experiments were done at large enough
flux so that one
suspects that equilibrium has not been reached yet. 
Finally, nucleation barriers might slow down the convergence to an equilibrium
state, trapping the system in metastable states \cite{jess}.

Our analysis indicates that for strain induced island formation there
is a critical strain, $\epsilon_1$, such that for any $\epsilon >
\epsilon_1$ {\it stable islands are possible}. A second important consequence is
that for any $\epsilon$, for {\it large enough coverage} {\it ripening
will occur}.  Thus in order to obtain stable islands, $H$ 
must {\it not} exceed the boundary of the
ripening phase. This result can provide an efficient test of our
predictions: in systems where ripening has been observed, we predict that
ripening can be avoided by choosing a smaller
coverage $H$.

The formation of the pseudomorphic wetting layer for small $H$ and
$\epsilon$ has been documented in various systems, being a general
feature of strained layer formation. Detailed measurements on
InAs/GaAs have shown that the transition from the FM to the SK$_1$
phase occurs at $H_{c_1} \simeq$1.7ML, independent of the deposition
rate \cite{gera}, indicating that its origin is thermodynamic rather
than dynamic \cite{daru2}.  Furthermore, recent investigations have measured the
strain dependence of $H_{c_1}$: results on GeSi grown on Si and AlInAs
grown on AlGaAs have indicated that the critical wetting layer
thickness decreases with increasing misfit \cite{abst}, in
agreement with the decreasing tendency of the $H_{c_1}$ phase boundary
(Fig. 1).

After the critical thickness has been reached, rapid formation of
uniform islands is observed \cite{mrs}.  Studies of InAs grown on GaAs
indicate that near $H_{c_1}$ the island density increases as
$\rho \sim (H-H_c(\epsilon ))^\gamma$ \cite{petr}, signaling a second
order phase transition in the system.  Furthermore, we find
that in the close vicinity of $H_{c_1}$ we have $\gamma =1$, i.e. the
island density increases linearly with ($H-H_{c_1}$).  However, for
large $(H-H_{c_1})$ the island-island interaction leads to sub-linear
increase in the density. Indeed, Miller {\it et al.} \cite{petr} found
that after stopping deposition the system had a transient regime,
after which it equilibrated. The {\it equilibrated} island density {\it increased
linearly} with the coverage, 
in agreement with our prediction $\gamma =1$. Furthermore, a linear expression provides 
an excellent fit near $H_{c_1}$ to the data of Ref. \cite{petr} as well.  

We find that unlike $\rho$, the equilibrium island size does not increase
continuously near $H_{c_1}$, but it
{\it jumps discontinuously} from zero to
$x_0(\epsilon , H_{c_1})$. This is again in agreement with the
experiments, since once islands form, they reach a well defined size
and small islands are rather rare \cite{mrs,will}.
Also, the experiments indicate that while an increasing $H$ does modify
the equilibrium island size, this change is not significant, but most
of the newly deposited material contributes to the formation of new
islands \cite{mrs}, again in agreement with slowly changing $x_0$ and rapidly
increasing $\rho$ in Fig. 2.

Finally, the phase diagram indicates that the stability of the islands
depends on the coverage: {\it independent of} $\epsilon$, {\it for
large} $H$ ripening should take place in the system. Indeed,  for
CdSe grown on ZnMnSe repeated AFM scans of the same sample made at 48
hour intervals indicate that some islands ripen at the expense of
others, and the overall island density decreases with time
\cite{xin}. However, the fact that the stable islands do not coexist
with the ripened ones, as expected in the R$_2$ and R$_3$ phases,
suggests that these experiments were performed either at the FM and
R$_1$ phase boundary, or dislocations relax the ripened
islands. The coexistence of stable and ripened islands is
documented for Ge grown on Si \cite{will}, that, together with the
evidence of a wetting film in this system, indicate that these
experiments are done at the border of the SK$_1$ and R$_2$ phases, and
consequently stable SK islands are allowed for smaller coverages,
assuming that dislocations are not the sole origin of the ripened islands. 

We have benefited from useful comments and discussions with J. K. Furdyna
and J. Tersoff.

\begin{figure}
\caption{Equilibrium phase diagram  in
function of the coverage $H$ and misfit $\epsilon$. The small panels
on the top and the bottom illustrate the morphology of the surface in
the six growth modes. The small empty islands indicate
the presence of stable islands, while the large shaded one refer to  ripened islands.
  The phases are separated by the
following phase boundary lines: $H_{c_1}(\epsilon )$: FM-R$_1$,
FM-SK$_1$; $H_{c_2}(\epsilon )$: SK$_1$-R$_2$; $H_{c_3}(\epsilon )$:
SK$_2$-SK$_1$; $H_{c_4}(\epsilon )$: VW-SK$_2$, VW-R$_3$. The
parameters used to obtain the phase diagram are $a=1$, $C=40E_0$,
$\Phi_{AA}=E_0$, $\Phi_{AB}=1.27E_0$, $g=0.7$, $p=4.9$, $\gamma =0.3$,
$b=10$.
 }
\end{figure}

\begin{figure}
\caption{(a) Wetting film thickness ($n_1$), island coverage
($n_2$) (top), island size ($x_0$), and island density ($\rho$)
(bottom) as a function of  $H$ for $\epsilon =0.08$.
 At $H_{c_1}$ there is a transition from the FM to the
SK$_1$ phase, the island size jumping discontinuously.  In the  R$_2$ phase, present
for $H > H_{c_2}$,  ripening takes place; 
(b) Same as (a) but  for $\epsilon =0.12$. 
 At $H_{c_4}$ there is a  transition  from the VW to the
SK$_2$ phase followed by the  SK$_1$ phase at H$_{c_3}$. And finally, at $H_{c_2}$ the system 
reaches the R$_2$ phase.}
\end{figure}

\end{multicols}

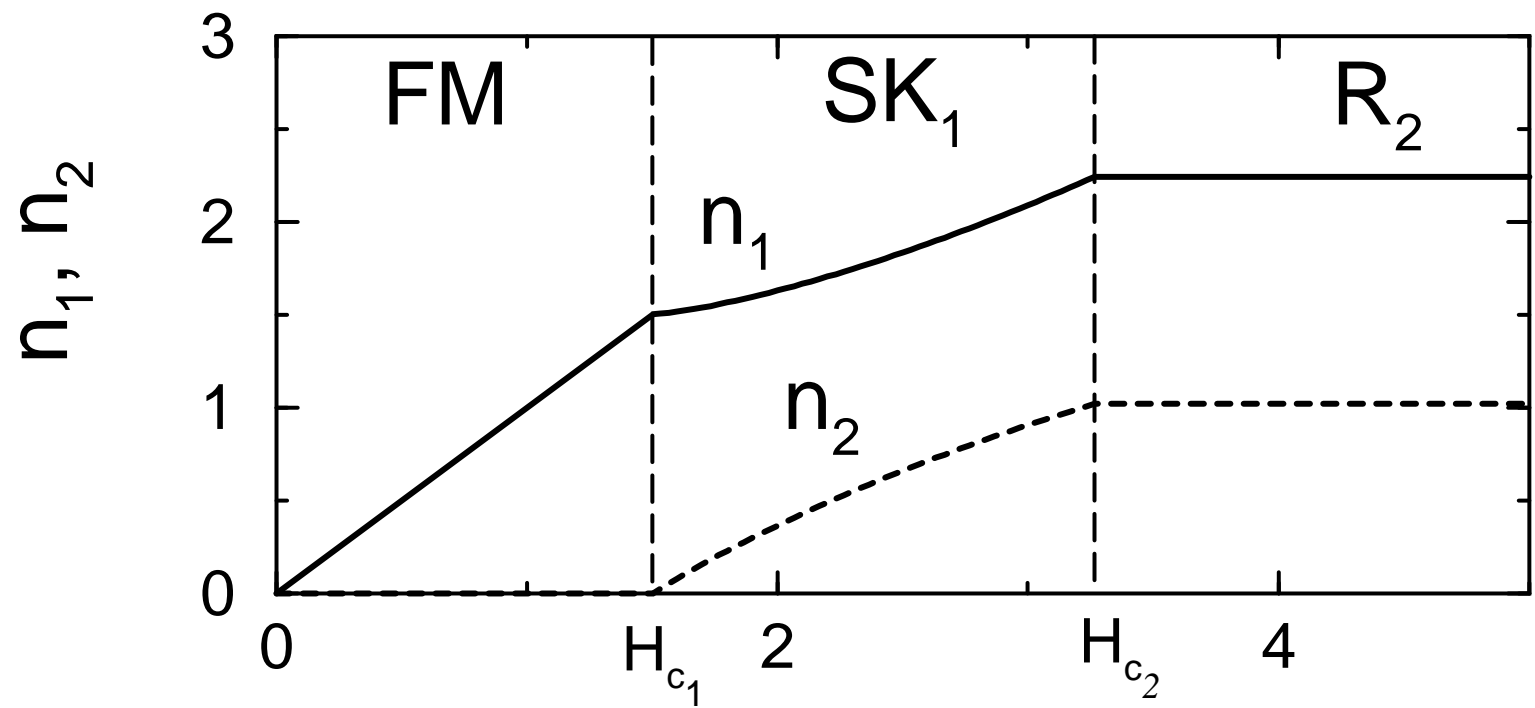
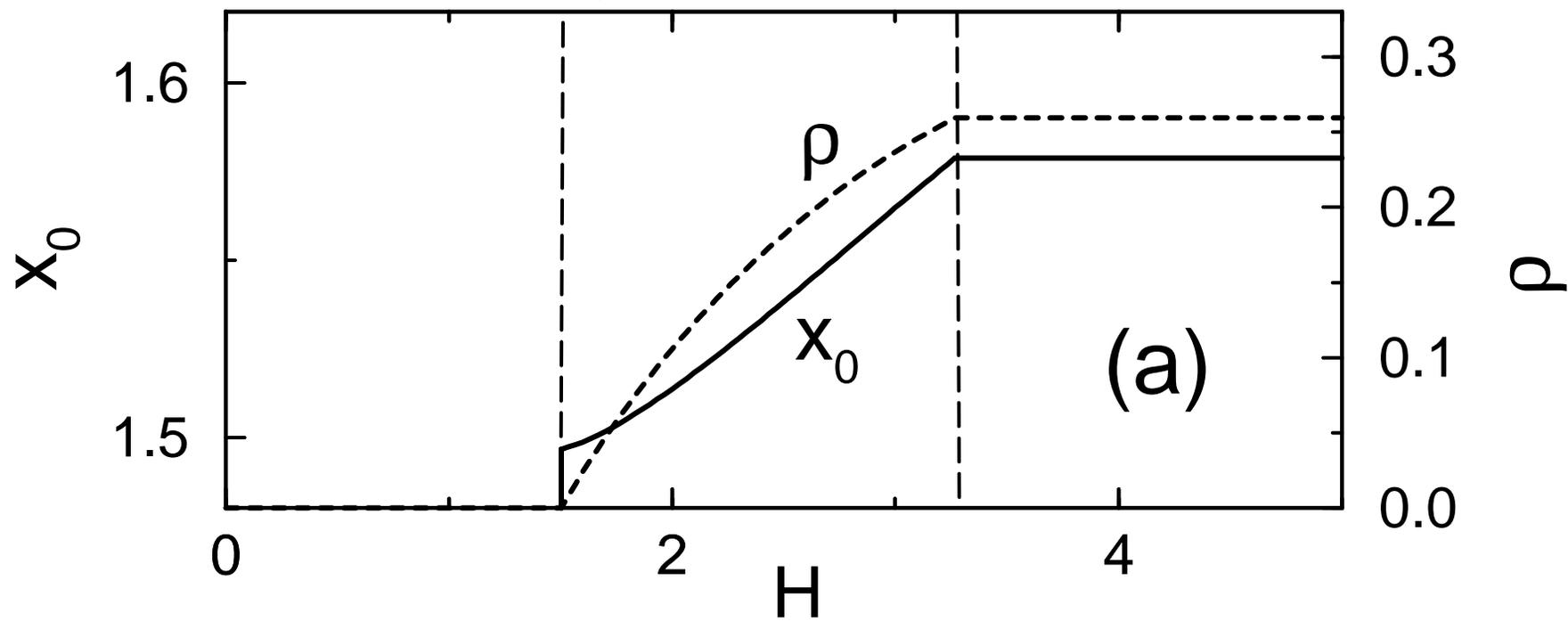
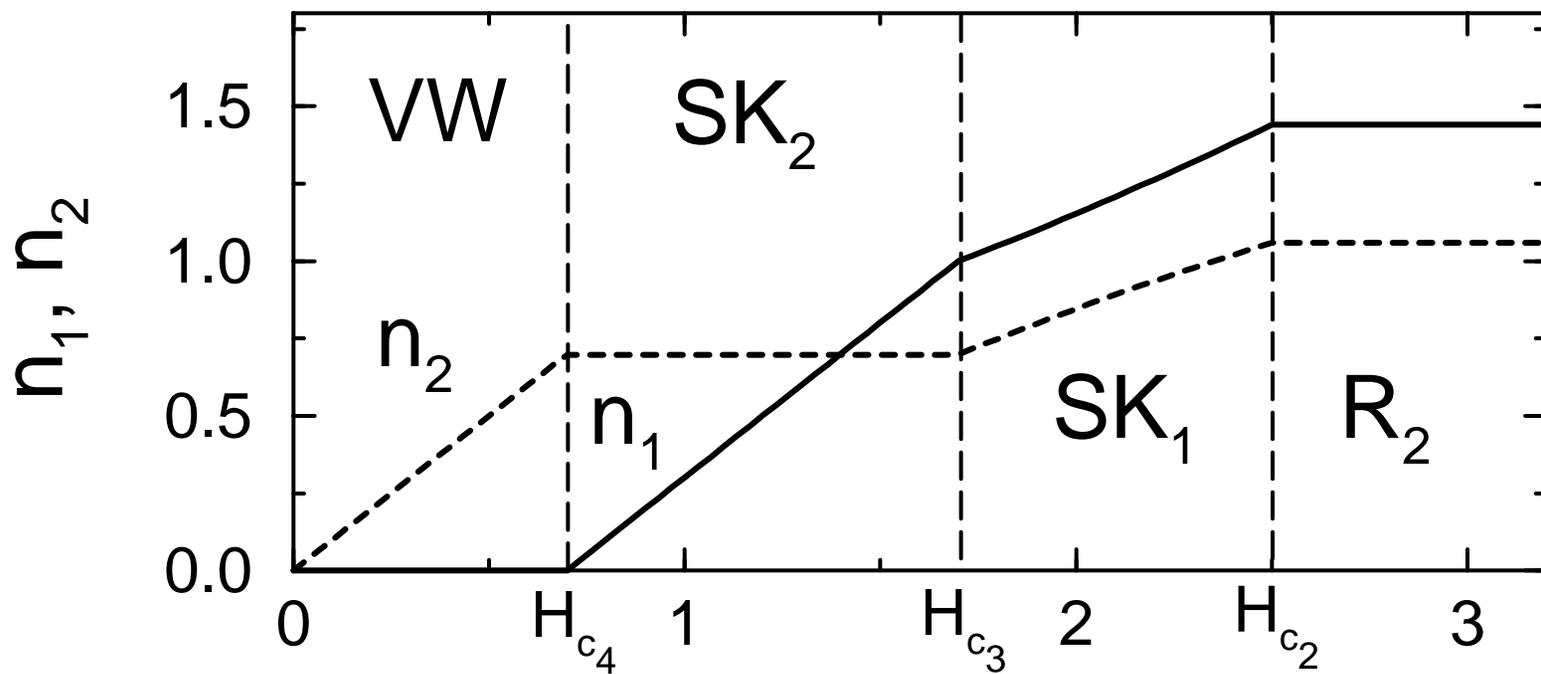
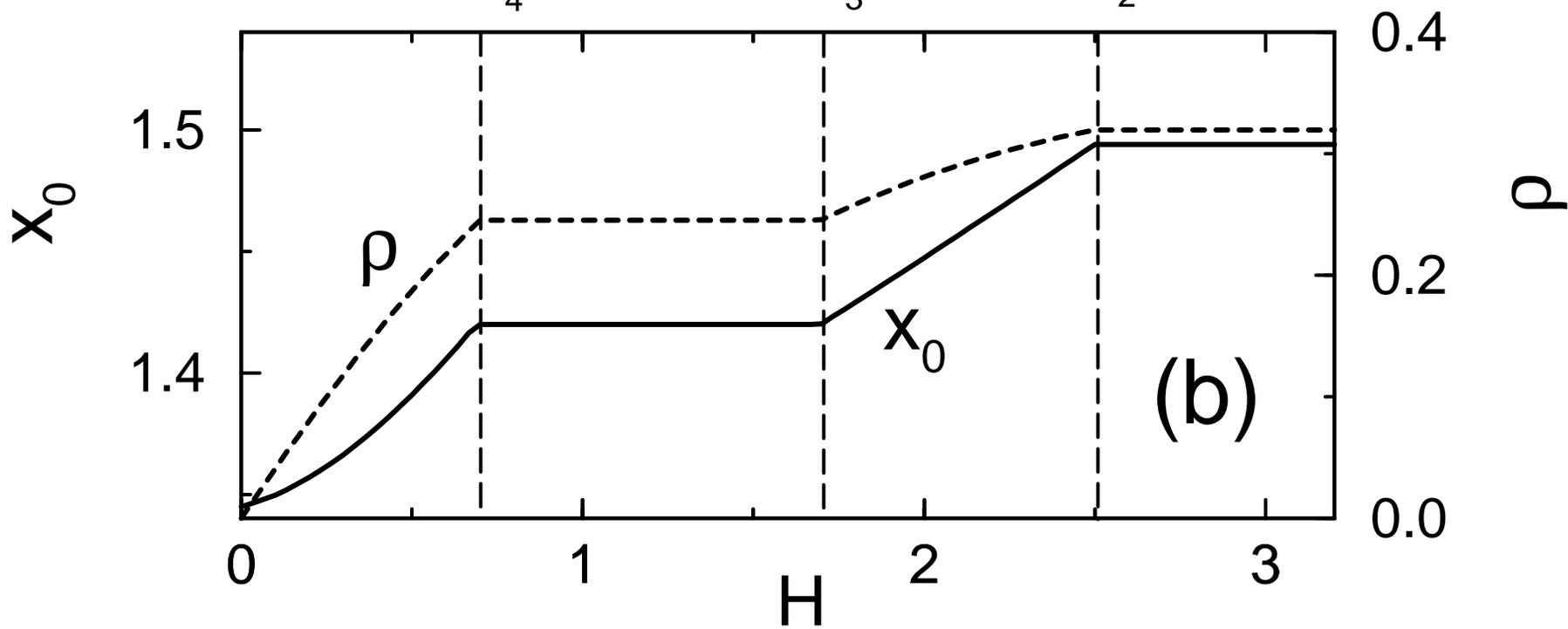
\end{document}